# COMPARATIVE STUDY BETWEEN LOW AND HIGH ENERGY SOHO/ERNE PROTONS IN SOLAR CYCLE 23

**Rositsa Miteva[1], Susan W. Samwel[2], Momchil Dechev[1]**

[1]*Institute of Astronomy with National Astronomical Observatory – Bulgarian Academy of Sciences*
[2]*National Research Institute of Astronomy and Geophysics – Helwan, Egypt*
*e-mail: rmiteva@nao-rozhen.org*

*Abstract:* *We present a comparison between low and high energy solar energetic protons (SEPs) as observed by the SOHO/ERNE high energy detector in solar cycle (SC) 23. Observed and corrected peak proton intensities are used in the analyses. The linear correlations are calculated between the proton intensity and the soft X-ray class of solar flares (SFs) on one side and the projected speed of coronal mass ejections (CMEs), on another. The energy trends of these correlations are compared with previous reports.*

### Introduction

Solar energetic protons (SEPs) are in situ observed fluxes of protons with energies ranging from MeVs up to GeVs (Desai and Giacalone, 2016). Numerous satellites (usually at L1, e.g. SOHO https://soho.nascom.nasa.gov/, geostationary orbits, e.g., GOES https://www.goes.noaa.gov/, and recently also closer to the Sun, e.g. PSP, http://parkersolarprobe.jhuapl.edu/, and Solar Orbiter, https://www.esa.int/Science_Exploration/Space_Science/Solar_Orbiter/) detect fluxes of electrons, protons and heavy ions continuously outflowing from the Sun.
SEPs can have numerous negative effects on technology, such as direct impact on the surface materials, electronics, optical systems, leading to discharges, ionization effects, software errors (e.g., Miteva et al. (2023) and the references therein). Another important aspect of the SEPs and their galactic counterparts is the risk of high radiation doses they may cause to astronauts. Based on data received from instruments aboard the ExoMars mission, Semkova et al. (2018) estimated that "During a cruise to Mars and back in declining solar activity the astronauts will accumulate at least 60% of the total career dose." Thus, the improved understanding on the SEP acceleration, transport and effects is crucial for their successful forecasting under the framework of space weather research (see e.g., https://spreadfast.astro.bas.bg/).

The debate on the solar origin: solar flares (SFs), Fletcher et al. (2011), vs. coronal mass ejections (CMEs), Webb and Howard (2012), either as the sole or the dominant particle accelerator seems to have reached a consensus, as nowadays both eruptive phenomena are recognized to be able to contribute to the particle fluxes (Trottet et al., 2015; Klein and Dalla, 2017).

In this report, we use proton data from the ERNE instrument aboard SOHO, Torsti et al. (1995) during solar cycle (SC) 23, namely in the period 1996–2009. Data only from the high energy detector (HED) will be considered. Similar data was used in our previous works on the topic, Miteva et al. (2018) and Miteva et al. (2020). The novelty in this report is the comparison of the Pearson correlation coefficients between the SEP peak intensities, both directly detected by the instrument and as corrected values with, on one side, the SF class, or with the projected CME speed, on another. For the correction of the SOHO/ERNE fluxes we will use the results reported in Miteva et al. (2020) and described in detail below. We focus on SC23, since we can directly compare our results with those reported by Dierckxsens et al. (2015). The latter is the only available report on the statistical preference between SFs and high-energy SEPs, and between CMEs and low-energy SEPs, based on GOES data. We examine these trends using an independent data source, namely SOHO/ERNE, within the same time period of SC23.

**Data and methodology**

The energy ranges (or channels) and their average value (in MeV) given in parenthesis of the HED instrument (https://export.srl.utu.fi/export_data_description.txt) are as follows:
- HED1: 13.8-16.9 (15.4)
- HED2: 16.9-22.4 (18.9)
- HED3: 20.8-28.0 (23.3)
- HED4: 25.9-32.2 (29.1)
- HED5: 32.2-40.5 (36.4)
- HED6: 40.5-53.5 (45.6)
- HED7: 50.8-67.3 (57.4)
- HED8: 63.8-80.2 (72.0)
- HED9: 80.2-101 (90.5)
- HED10: 101-131 (108)

For the purpose of this report, we use the lowest, HED1, as well as the highest energy channel, HED10, in both observed and corrected fluxes.

We use the results from our previous work, Miteva et al. (2020), on the correction of SOHO/ERNE peak proton intensities (denoted as $J_{p,l,corr}$ and $J_{p,h,corr}$) using the Wind/EPACT proton fluxes as calibration. We have no Wind/EPACT channels at ~15 MeV or at ~108 MeV for which to perform the cross-correlation procedure. Thus, after performing several tests we decided to increase only the limit at the low energy Wind/EPACT flux for which we apply the correction, namely:
- when the Wind/EPACT low energy flux >= 8 DPFU:
$J_{p,l,corr}$ (ERNE) = -0.32+0.89*$J_{p,l}$ (EPACT)
- when the Wind/EPACT high energy flux >= 1 DPFU:
$J_{p,h,corr}$ (ERNE) = -0.37+0.81*$J_{p,h}$ (EPACT)

The differential proton flux unit (DPFU) is measured in protons per (cm^2 sr s MeV).
In summary, the above relationships which are used to calculate the correction fluxes should be considered as approximations only, since the fitting procedure is done for Wind/EPACT protons with 25 MeV and 50 MeV energies, respectively, instead of 10 and 100 MeV. The corrected fluxes, denoted with filled black color, are shown in Fig. 1 and are obtained by performing the above correction equations on the fluxes denoted with empty black circles. The events denoted with filled magenta-colored circles (denoting the original proton flux, not corrected) and filled black circles (denoting the corrected fluxes) are finally used in the associations below. The empty circles denote the detected proton flux subject to saturation effects, which is clearly evident for the low-energy sample (Fig. 1 left). Some overcorrection is recognized for the high energy flux and the results should be used with caution or regarded as an upper limit only. In total, the corrected events are 27 for the lowest and 24 for the highest energy channel. Finally, the correlations results based on the observed fluxes should be considered as lower limits for the obtained statistical trends.

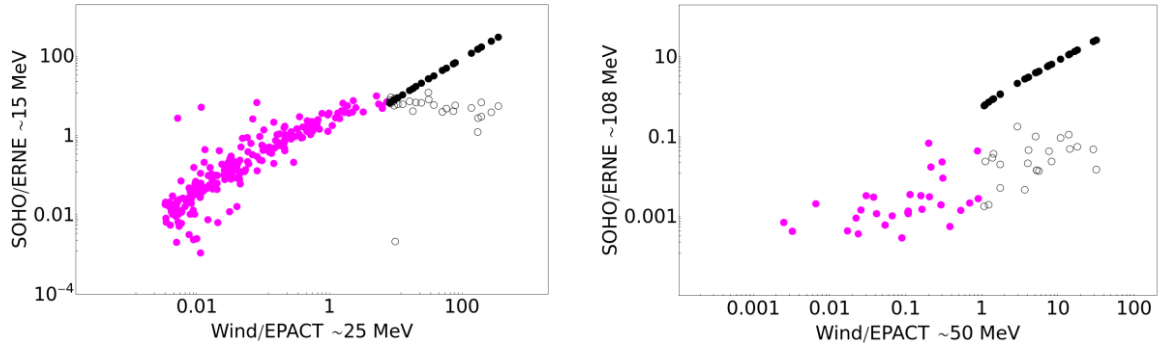

Fig. 1. Scatter plots between the proton fluxes using SOHO/ERNE and Wind/EPACT low (left) and high (right plot) energy channels. For the color code used see text.

### Results

For this short report, we selected all 444 SEPs recorded in the HED1 channel. This sample of protons events could be associated with 297 SFs (or 67% of the protons have a SF-origin) and 342 with CMEs (77% of the protons have a CME-origin). Only 55 SEPs could be identified in the HED10 channel. The data for the CME speed are taken from https://cdaw.gsfc.nasa.gov/CME_list/ whereas for the SF flux are from https://www.ncei.noaa.gov/products/space-weather/legacy-data/solar-flares.

In Table 1 are summarized all Pearson correlation coefficients between the SOHO/ERNE observed and corrected (for low, 15 MeV and high, 108 MeV) SEP fluxes with SF class (denoted with ISXR) and CME speed (denoted with VCME). The later correlations are done as log10-linear and log10-log10. We add the results from Dierckxsens et al. (2015) for the following energy channels: 5.00–7.23 MeV (6 MeV); 10.46–15.12 MeV (12.79) and 95.64–138.3 MeV (117).

Table. 1. Pearson correlation coefficients between the observed and corrected values of SEP peak flux with SF class and CME linear speed for SC23. In parentheses is given the sample size used for each calculation.

|  | log10 Jp vs. log10 ISXR | log10 Jp vs. linear VCME | log10 Jp vs. log10 VCME |
|---|---|---|---|
| **Low energy channel: 15 MeV** | | | |
| HED1 onserved | 0.49±0.05 (297) | 0.55±0.04 (342) | 0.52±0.04 (342) |
| HED1 corrected | 0.50±0.05 (297) | 0.57±0.04 (342) | 0.53±0.04 (342) |
| Dierckxsens et al. | (6 MeV: 0.38±0.09) 13 MeV: 0.52±0.08 | (6 MeV: 0.59±0.06) 13 MeV: 0.57±0.07 | - |
| **High Energy channel: 108 MeV** | | | |
| HED10 observed | 0.26±0.12 (52) | 0.36±0.11 (52) | 0.42±0.10 (52) |
| HED10 corrected | *0.38±0.12 (52)* | *0.50±0.11 (52)* | *0.52±0.09 (52)* |
| Dierckxsens et al. | 117 MeV: 0.59±0.07 | 117 MeV: 0.32±0.10 | - |

No statistical differences are obtained between the correlation coefficients using observed and corrected SEP fluxes for the case of low energy protons. Concerning the high energy channel, the differences are larger, however not statistically significant, as the uncertainty is 9–12%. In addition, the corrected high energy SEP fluxes are subject to overcorrection (given with italic font in Table 1).

The Pearson correlation coefficients between the low-energy SEPs (15 MeV) and SF class are reported both for observed and corrected values in the second column and have the same values (~0.5). The log10-linear correlations obtained with the CME speed are in the range 0.55–0.57, whereas the log10-log10 values are ~0.52, however the differences are not statistically significant. The results obtained by us are consistent with the reported ones by Dierckxsens et al. (2015) based on 13 MeV protons.

Concerning the high energy protons, a slight increase is noticed for the high-energy SEP trends, when comparing the correlations between SEPs and SF (0.26) vs. those with CMEs (0.36). Even larger values are obtained when using the corrected fluxes. Since the correction for the high-energy SEP fluxes is dubious, the results should be taken with caution. In summary, based on the observed SEP fluxes from SOHO/ERNE, we obtained similar results for the correlations with the CMEs compared to Dierckxsens et al. (2015). Regarding the correlation of high energy protons with SF class (0.26±0.12), opposite trends are found by us compared to those obtained by Dierckxsens et al. (2015), namely 0.59±0.07 and the difference is statistically significant.

## Summary

In our previous work, Miteva et al. (2018), their Fig. 5 (left), we showed the declining energy trends (based on detected peak proton intensities with SFs and CMEs) of HED2, 4, 6, 7, 10 channels in SC23 (using data by the end 2016). The results in this study based on HED1, 10 confirm these previous trends also using corrected SOHO/ERNE proton fluxes. The corrected SEP fluxes tend to show larger values for the correlation coefficient compared to those derived by the observed fluxes, however these are not statistically different from the correlations based on observed data. In summary, the correlations using SOHO/ERNE SEP data do not confirm only the reported trend by Dierckxsens et al. (2015) for the relationship between high-energy protons and SFs. Further extension of this study covering also SC24 is underway and will be reported elsewhere. The SOHO/ERNE SEP catalog at several energy channels together with the associated solar origin will be made freely accessible at a dedicated web-site: https://catalogs.astro.bas.bg/.

## Acknowledgements

The results in this study are part of: (1) The inter-academy project 'On space weather effects at near Earth environment - from remote observations and in situ particle forecasting to impacts on satellites' funded by the Bulgarian Academy of Sciences IC-EG/08/2022-2024 and Egyptian Academy of Scientific Research and Technology (ASRT)/NRIAG (ASRT/BAS/2022-2023/10116; (2) The Bulgarian-Serbian bilateral project 'Active Events On The Sun. Catalogs Of Proton Events And Electron Signatures In X-Ray, UV And Radio diapason. Influence of Collisions on Optical Properties of Dense Hydrogen Plasma' funded by the Bulgarian Academy of Sciences and (3) the SCOSTEP/PRESTO 2020 grant 'On the relationship between major space weather phenomena in solar cycles 23 and 24'.